\documentclass[a4paper,superscriptaddress,showpacs,twocolumn,prl]{revtex4-1}
\usepackage{amsfonts}
\usepackage{amsmath}
\usepackage{amssymb}
\usepackage{graphicx}
\usepackage{dsfont}

\begin{document}

\title{Wave-particle duality in an environment with arbitrary white noise}

\author{J. G. Filgueiras}
\email{jgfilg@cbpf.br}
\affiliation{Centro Brasileiro de Pesquisas F\'{\i}sicas, Rua Dr. Xavier Sigaud 150, 22290-180 Rio de Janeiro, Rio de Janeiro, Brazil}

\author{R. S. Sarthour}
\affiliation{Centro Brasileiro de Pesquisas F\'{\i}sicas, Rua Dr. Xavier Sigaud 150, 22290-180 Rio de Janeiro, Rio de Janeiro, Brazil}

\author{A. M. Souza}
\affiliation{Centro Brasileiro de Pesquisas F\'{\i}sicas, Rua Dr. Xavier Sigaud 150, 22290-180 Rio de Janeiro, Rio de Janeiro, Brazil}

\author{I. S. Oliveira}
\affiliation{Centro Brasileiro de Pesquisas F\'{\i}sicas, Rua Dr. Xavier Sigaud 150, 22290-180 Rio de Janeiro, Rio de Janeiro, Brazil}

\author{R. M. Serra}
\affiliation{Centro de Ci\^{e}ncias Naturais e Humanas, Universidade Federal do ABC, R.  Santa Ad\'{e}lia 166, 09210-170 Santo Andr\'{e}, S\~{a}o Paulo, Brazil}

\author{L. C. C\'{e}leri}
\email{lucas@chibebe.org}
\affiliation{Centro de Ci\^{e}ncias Naturais e Humanas, Universidade Federal do ABC, R.  Santa Ad\'{e}lia 166, 09210-170 Santo Andr\'{e}, S\~{a}o Paulo, Brazil}

\begin{abstract}
The development of quantum technologies depends on investigating of the behavior of quantum systems in noisy environments, since complete isolation from its environment is impossible to achieve. In this paper we show that a wave-particle duality experiment performed in a system with an arbitrarily white noise level cannot be explained in classical terms, using hidden-variables models. In the light of our results, we analyze recent optical and NMR experiments and show that a loophole on non-locality is not fundamental.
\end{abstract}


\maketitle


The recent developments of new quantum technologies enable the experimental realization of various experiments, which can be used to test some very basic aspects of quantum mechanics and its application in quantum computation and communication. To cite a few examples, non-local aspects have been tested via Bell-like inequalities \cite{bell}, the violation of macroscopic realism through Legget-Garg inequalities \cite{garg} and quantum contextuality verified in \cite{cabello}. More recently, the concept of weak measurements revealed the need to revisit Heisenberg's uncertainty principle \cite{ozawa}, which has been experimentally verified using neutron interferometry and quantum optics devices \cite{unct}. Furthermore, a fundamental concept of quantum mechanics, the quantum complementarity principle, has also been tested recently \cite{DSE}. This principle is related to the wave-like (WL) behavior, revealed by the appearance of interference patterns, and particle-like (PL) behavior. It states that WL and PL behaviors are complementary and mutually excludent. This fact, which has been largely verified in different experiments \cite{DSE}, means that in order to observe these phenomena, we need two distinct experimental arrangements. 

\begin{figure}
\includegraphics[scale=0.3]{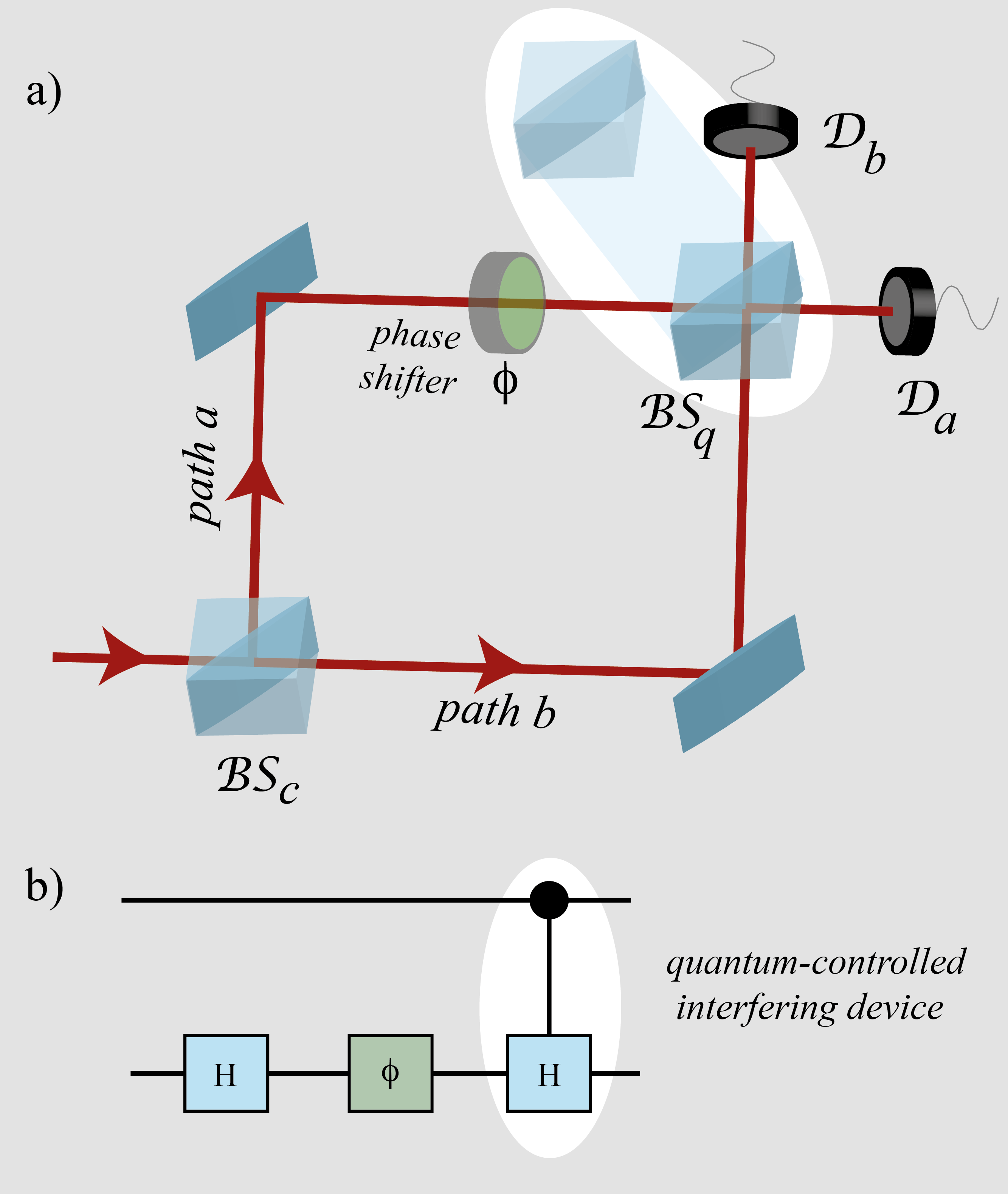}
\caption{(a) Schematic diagram of the Mach-Zehnder interferometer with a \textit{quantum} beam splitter $BS_{2}$. (b) The quantum circuit that describes the evolution of the ancilla and the photon in the interferometer. The ancilla is the qubit in the first line of the circuit, while the qubit inside the interferometer is in the second line. The state of the ancilla is given by $\cos\alpha\vert 0\rangle + \sin\alpha\vert 1\rangle$. $H$ is the Hadamard gate and $\varphi$ is a gate that creates the phase difference $\varphi$ between the paths $a$ and $b$. The interferometer is closed for $\alpha = \pi/2$ and open for $\alpha = 0$. For any other value, $0<\alpha<\pi/2$, the interferometer is in a coherent superposition of being closed and open.}
\label{Fig1}
\end{figure}

For example, in a Mach-Zehnder interferometer the interference pattern appears in the detectors $\mathcal{D}_{a}$ and $\mathcal{D}_{b}$ when the interferometer is closed, i.e., when the second beam splitter (see figure \ref{Fig1} (a)) is present. In order to explain this phenomenon, it is necessary to assume that the photon has travelled through both paths of the interferometer simultaneously. In the other manner, if the second beam splitter is absent, the experiment reveals a which-path information. In the language of the complementarity principle, if we want to observe the wave aspect of the photon, we must consider the closed interferometer (with $\mathcal{BS}_{q}$ present), whereas to observe the particle nature of the photon, we must remove the second beam splitter $\mathcal{BS}_{q}$ (the open interferometer). This means that these two different experimental arrangements are complementary. Either choice determines the statistics of the results beforehand by the experimenter's decision.

This fact led Wheeler to formulate his famous delayed-choice \emph{gedanken} experiment \cite{Wheeler}. Wheeler's main idea was based on the question: Is it possible for the photon to ``know'' the experimenter's choice and then, behave accordingly? To answer this question Wheeler proposed to make the choice (between the open or closed interferometer) only after the photon passes through the first beam splitter \cite{Wheeler}. This  experiment has been recently performed in an optical setup and the complementarity principle was elegantly confirmed \cite{DCC}, showing that there is no difference between the normal and the delayed versions of the experiment. This is a classical experiment, in the sense that the interferometer has only two states, open \emph{or} closed. 

Recently, the quantum extension of the delayed choice experiment has been proposed \cite{Terno} and experimentally verified for spins \cite{Auccaise, Mahesh}, as well as for photons \cite{Popescu, Kaiser, Guo}. The extension was to consider the second beam splitter in a coherent superposition of being present and absent. The second beam splitter, $\mathcal{BS}_{q}$, is now controlled by a quantum device, referred to as the ancilla system, which allows the beam splitter to be in a superposition of being present and absent. The simple quantum circuit shown in figure \ref{Fig1} (b) describes the evolution of the system through the interferometer. If the initial state of the ancilla-system is given by
\begin{equation}
\vert\psi\rangle_{\mathcal{SA}}^{I} = \vert 0\rangle_{\mathcal{S}}\otimes\left[\cos\alpha\vert 0\rangle _{\mathcal{A}} + \sin\alpha\vert 1\rangle _{\mathcal{A}}\right],
\label{initial_state}
\end{equation}
then the final system-ancilla state, after the action of the second beam splitter, is given by \cite{Terno, Auccaise}
\begin{equation}
\vert \psi \rangle= \cos\alpha\vert p\rangle_{\mathcal{S}}\vert 0\rangle_{\mathcal{A}}+ \sin\alpha\vert w\rangle_{\mathcal{S}}\vert 1\rangle_{\mathcal{A}}.
\label{final_state}
\end{equation}
In this equation, $\vert p\rangle_{\mathcal{S}} = \left(\vert 0\rangle_{\mathcal{S}}+ e^{i\phi}\vert 1\rangle_{\mathcal{S}}\right)/\sqrt{2}$ denotes the PL behavior of the photon while $\vert w\rangle_{\mathcal{S}} = e^{i\phi/2}\left(\cos\left(\phi/2\right)\vert 0\rangle_{\mathcal{S}} - i\sin\left(\phi/2\right)\vert 1\rangle_{\mathcal{S}}\right)$ accounts for its WL behavior. The reason for these identifications is that $\vert p\rangle_{\mathcal{S}}$ and $\vert w\rangle_{\mathcal{S}}$ state vectors result in particle and wave statistics, respectively.

The introduction of the quantum device has two major consequences. The first one is that the experimenter is able to choose which behavior (WL or PL) will be observed \emph{after} the measurement. Both behaviors are obtained by correlating the experimental data for the ancilla and for the system measurements. The complementarity is not in the experimental arrangements, since only one is used in the experiment, but lies in the correlations between the experimental data \cite{Terno, Auccaise, Mahesh, Popescu, Kaiser, Guo}. Such property gives the possibility of a smooth transition between the WL and PL behaviors of the system. The second one, and maybe the most important aspect, was the demonstration that, for the pure state case, there is no local hidden-variable model reproducing the joint probability distribution obtained from the state described in Eq.(\ref{final_state}). This means that the wave or particle behaviors of the photon are not \emph{realistic} properties, i.e., intrinsic properties of the system, which are independent of the experimental arrangement \cite{Terno}. 

Quantum phenomena are usually valuable resources for computation and communication tasks \cite{nielsen}. From a practical point of view it is important to know to what extent noise affects quantum coherence. In this paper we analyze how the presence of white noise affects the quantum delayed choice experiment and show that, like in the pure state case, the wave-particle duality cannot be explained in classical terms. Another important feature concerns the role of entanglement in this experiment, since the state given by Eq. (\ref{final_state}) is entangled for all values of $\alpha$ except for the two extremal values. 

To start, let us define the notation. Following \cite{Auccaise}, states $\vert 1\rangle_{\mathcal{S}}$ and $\vert 0\rangle_{\mathcal{S}}$ will be used to label the interferometric paths $a$ and $b$, respectively. The transformation employed by the second beam splitter is coherently controlled by the ancillary system, in such a way that, if it is in the state $\vert 0\rangle_{\mathcal{A}}$ the second beam splitter is absent, meaning that the interferometer is open, while if its state is $\vert 1\rangle _{\mathcal{A}}$, $\mathcal{BS}_{q}$ is present and then, we have a closed interferometer. As $\mathcal{BS}_{q}$ is now a quantum system, its state is not limited to be present or absent, it can be any superposition between $\vert 0\rangle_{\mathcal{A}}$ and $\vert 1\rangle_{\mathcal{A}}$, meaning that the interferometer can be cast in an arbitrary superposition of being open and closed \cite{Terno}.

In the usual computational basis $\left\{00,01,10,11\right\}$ in the ${\mathcal{S}\otimes\mathcal{A}}$ space, the final joint probability distribution of the system and the ancilla is given by \cite{Terno}
\begin{eqnarray}
P\left(S,A\right) &= \left[\frac{1}{2}\cos^{2}\alpha,\sin^{2}\alpha\cos^{2}\frac{\phi}{2}, \right.\nonumber \\
& \left.\frac{1}{2}\cos^{2}\alpha,\sin^{2}\alpha\sin^{2}\frac{\phi}{2}\right].
\label{statpuro}
\end{eqnarray}
with $S$ $\in$ $\mathcal{S}$ and $A$ $\in$ $\mathcal{A}$ representing the measurement outcomes in the computational basis.

In order to show that the wave-particle duality cannot be described by a local hidden-variable model in the presence of white noise, we will consider the Werner state \cite{note} given by
\begin{equation}
\rho_{\mathcal{SA}} = \eta\mathds{1}_{\mathcal{SA}} + \varepsilon\sigma_{\mathcal{SA}} ,
\label{state}
\end{equation}
\par\noindent
in which $\eta = (1 - \varepsilon)/4$, with $\varepsilon$ quantifying the purity of the state and $\sigma_{\mathcal{SA}}$ is the pure density matrix associated with the state given by Eq. (\ref{initial_state}). When $\varepsilon = 1$ we recover the pure state case, with probability distribution given by Eq. (\ref{statpuro}). However, for any other value of $\varepsilon$, we have a mixed state as the input for the experiment. Remembering that the whole interferometer is a unitary operation, the final joint probability distribution is given by
\begin{eqnarray}
P_{\varepsilon}\left(S,A\right) &=& \mbox{Tr}\left[U_{I}\rho_{\mathcal{SA}}U_{I}^{\dagger}\hat{P}_{\mathcal{SA}}\right] \nonumber \\
&=&\eta + \varepsilon P\left(S,A\right),
\label{statmisto}
\end{eqnarray}
with $\hat{P}_{\mathcal{SA}} = \vert S\rangle\otimes\vert A\rangle\langle S\vert\otimes\langle A\vert \equiv \vert SA\rangle\langle SA\vert$ being the projector operator, in the computational basis, over the $\mathcal{SA}$ space, $P\left(S,A\right)$ is given by Eq. (\ref{statpuro}) and $U_{I}$ is the unitary operator describing the evolution of $\rho_{\mathcal{SA}}$ through the interferometer.

In what follows, we prove that, independent of the value of $\varepsilon$, there is no local hidden-variable model that describes the statistics given by Eq. (\ref{statmisto}), as observed in two recent experiments \cite{Auccaise, Mahesh}.

Following the proposal of Ref. \cite{Terno} we assume that each photon presents WL or PL behaviors, with certain probabilities. Such properties are determined only by the source of the photons, and completely independent of any possible measurement scheme. In other words, we assume that there is some hidden-variable $\lambda$ that encodes the information about the particle ($\lambda = p$) or wave ($\lambda = w$) character of the photon. Here, $p$ and $w$ may represent distinct values of a certain variable or a set of values of a set of variables. In this model, the experimental probability distribution $P_{\varepsilon}\left(S,A\right)$ is the marginal distribution involving $\lambda$, i.e. $P_{\varepsilon}\left(S,A\right) = \sum_{\lambda}P_{\varepsilon}\left(S,A,\lambda\right)$, which, using Bayes' rule, can be rewritten in the more convenient form
\begin{equation}
P_{\varepsilon}\left(S,A\right) = \sum_{\lambda}P_{\varepsilon}\left(S\vert A,\lambda\right)P_{\varepsilon}\left(A\vert\lambda\right)P_{\varepsilon}\left(\lambda\right),
\label{eq1}
\end{equation}
with $P\left(X\vert Y\right)$ as the conditional probability of $X$ given $Y$.

The next step is the definition of all unknown distributions needed to compute $P_{\varepsilon}\left(S,A,\lambda\right)$:

\begin{itemize}
\item The probability distribution of the hidden-variable
\begin{equation}
P_{\varepsilon}\left(\lambda\right) = \left[a,1-a\right],
\label{distlambda}
\end{equation}
where $a$ is the probability of $\lambda$ to assume the value $p$ and $(1-a)$ for the value $w$.

\item The conditional probabilities
\begin{eqnarray}
&P_{\varepsilon}\left(A\vert\lambda = p\right) = \left[b,1-b\right],\nonumber\\
&P_{\varepsilon}\left(A\vert\lambda = w\right) = \left[c,1-c\right].
\label{distcondA}
\end{eqnarray}

\item The behavior of a WL photon in an open interferometer ($\mathcal{A} = 0$) is unknown. Therefore
\begin{equation}
P_{\varepsilon}\left(S\vert A = 0, \lambda = w\right) = \left[d,1-d\right].
\label{distS0W}
\end{equation}
The same occurs for the behavior of a PL photon in a closed interferometer ($\mathcal{A} = 1$)
\begin{equation}
P_{\varepsilon}\left(S\vert A = 1, \lambda = p\right) = \left[e,1-e\right].
\label{distS1P}
\end{equation}
\end{itemize}

All other distributions can be computed directly from $P_{\varepsilon}\left(S,A,\lambda\right)$ given in Eq. (\ref{statmisto}), as marginal distributions.

\begin{itemize}
\item  For the ancilla we have $P_{\varepsilon}\left(A\right) = \sum_{S}P_{\varepsilon}\left(S,A\right)$ leading to
\begin{eqnarray}
&P_{\varepsilon}\left(A = 0\right) = 2\eta + \varepsilon\cos^{2}\alpha \nonumber\\
&P_{\varepsilon}\left(A = 1\right) = 2\eta + \varepsilon\sin^{2}\alpha.
\label{distancilla}
\end{eqnarray}
upon the constrains
\begin{equation}
P_{\varepsilon}\left(A\right) = \sum_{\lambda}P_{\varepsilon}\left(A\vert\lambda\right) P_{\varepsilon}\left(\lambda\right)
\label{vinculo}
\end{equation}

\item The behavior of a PL photon in an open interferometer ($A = 0$ and $\lambda = p$) and of a WL one in a closed interferometer ($A = 1$ and $\lambda = w$) are
\begin{equation}
P_{\varepsilon}\left(S\vert A = 0, \lambda = p\right) = \left[\frac{1}{2},\frac{1}{2}\right].
\label{distS0P}
\end{equation}
and
\begin{eqnarray}
P_{\varepsilon}\left(S\vert A = 1, \lambda = w\right) &=& \frac{1}{p_{1}}\left[\eta + \varepsilon\cos^{2}\frac{\phi}{2}\sin^{2}\alpha;\right.\nonumber\\
&&\left. \eta + \varepsilon\sin^{2}\frac{\phi}{2}\sin^{2}\alpha\right],
\label{distS1W}
\end{eqnarray}
respectively. We remark that, in this last equation, we always have $p_{1} > 0$ for any positive $\varepsilon$, since $\alpha$ must be real.
\end{itemize}
In this last equation, $p_{1} \equiv p_{A=1} = 2\eta + \varepsilon\sin^{2}\alpha$ is the probability of measuring $A$ in state $\vert 1\rangle_{\mathcal{A}}$.

By means of Eq. (\ref{eq1}), together with the above computed distributions, we obtain the following set of equations
\begin{equation}
c\left(1-a\right)\left(d-\frac{1}{2}\right) = 0,
\end{equation}
\begin{equation}
a\left(1-b\right)\left(e-\beta\right) = 0,
\end{equation}
\begin{equation}
ab + c\left(1-a\right) - p_{0} = 0,
\end{equation}
where we have defined
\begin{equation}
\beta = \frac{\eta + \varepsilon\cos^{2}\left(\phi/2\right)\sin^{2}\alpha}{1 - p_{0}}
\end{equation}
and $p_{0} = 1 - p_{1}$.

If this set of equations does not present any valid solution, we are led to conclude that there is no hidden-variable model that describes the observed probability, since Eq. (\ref{eq1}) is not satisfied.

As $\beta$ and $\alpha$ are arbitrary we have to disregard the trivial solutions $(c=0,a=0)$ and $(a=1,b=1)$, because they would imply $p_{0} = 0$ and $p_{1} = 0$, respectively.

The solution $d = 1/2$ implies that, in an open interferometer, WL photons will present PL statistics
\begin{equation}
P_{\varepsilon}\left(S\vert A = 0, \lambda = w\right) = \left[\frac{1}{2},\frac{1}{2}\right],
\end{equation}
which is clearly not acceptable. A comment about high levels of noise is necessary here. In Eq. (\ref{distS1W}) we have WL photons presenting particle statistics in a closed interferometer for the case $\varepsilon = 0$. However, only in this case, $\rho_{\mathcal{SA}}$ is equal do the identity and as the interferometer is a unitary operation, nothing happens to the system. We will return to this point in the discussions.

The solution $e = \beta$ implies that
\begin{equation}
P_{\varepsilon}\left(S\vert A = 1, \lambda = p\right) = \left[\beta,1 - \beta\right].
\end{equation}
As $\beta$ is $\phi$ dependent, \emph{for any $\varepsilon > 0$}, this solution means that PL photons, in a closed interferometer, will exhibit a WL behavior (see Eq. (\ref{distS1W})), which is also not acceptable.

All of these solutions must be disregarded as the photon would exhibit inconsistent behaviors, contradicting the hypothesis of the reality of wave and particle aspects.

The last solution, $c = 0$ and $b = 1$, leads us to $a = p_{0}$. This means that the photons are randomly distributed as $P_{\varepsilon}(\lambda) = \left[p_{0},1-p_{0}\right]$, which is the same distribution as $P_{\varepsilon}(A)$. The hidden-variable $\lambda$ and the ancilla are perfectly correlated, which implies that the hidden-variable would be determined by the experimenter's choice for the value of $\alpha$. Therefore, due to Occam's razor, this solution must also be disregarded, because if $\lambda$ completely determines the value of the ancilla, then it cannot be determined by the experimenter's choice for the control parameter $\alpha$ in the preparation of the ancilla.

This result implies that, \emph{for any $\varepsilon > 0$}, the probability distribution given in Eq. (\ref{statmisto}) cannot be explained by a local hidden-variable model in which particle and wave are physical characteristics of quantum systems.

In the original proposal of the delayed-choice experiment, the second beam splitter is a classically controlled device \cite{Wheeler}. After the photon passes through the first beam splitter, an ancilla quantum system is prepared and measured. Depending on the outcome of the measurement, the choice between an open or closed interferometer is made. Note that the time ordering of the measurements (ancilla, then photon) is capital in this case. In the quantum version \cite{Terno}, the classical control is replaced by a quantum device, which puts the second beam splitter in a coherent superposition of being present and absent. This superposition leads to a correlation between the outcome of the ancilla and the behavior of the system in the interferometer. Instead of measuring the ancilla, the photon coherently interacts with it and then both systems are measured. The introduction of the quantum device causes a freedom of choice in the time ordering of the measurements, which is not present in the classical case. Here, the ancilla could be measured \emph{after} the photon.
\par\noindent
\textit{Experimental implications.} In the pure state case, the state given by Eq. (\ref{final_state}) is entangled for  
$0$ $< \alpha < \frac{\pi}{2}$. The entanglement between the system and ancilla, detected by the violation of Bell's inequality, was used in the works \cite{Popescu,Kaiser} as a proof of the quantum behavior of the interferometer. On the other hand, we considered here the state described by Eq. (\ref{final_state}) in the presence of white noise, i.e., the Werner state given in Eq.(\ref{state}). For $\varepsilon < 1/3$ this state is separable (non-entangled) for any $\alpha$ \cite{Braunstein}. We showed that the proof given in \cite{Terno} can also be extended to Werner states for any $\varepsilon > 0$. This result indicates that the system-ancilla entanglement is not the main feature for the absence of a classical model reproducing the quantum mechanical distribution. Furthermore, this indicates that the entanglement is not related to the correlation between the ancilla and the WL or PL of the system, since this correlation is robust against an \emph{arbitrary} level of white noise.
 
The experiments reported in Refs. \cite{Auccaise,Mahesh} were performed in a molecular quantum simulator in which there is a high level of noise. Nuclear spins are used to encode the interferometric system and the control ancilla. It is a well-known fact that in such systems it is not possible to create entanglement \cite{Braunstein}, but it is possible to employ non-factorizable gates to generate quantum correlations (of separable states) between the system and the ancilla, even in the presence of high levels of noise ($\varepsilon \ll 1$) \cite{discord}. The possibility to perform such kind of gates allows implementing of the quantum version of the delayed choice experiment. It is well known that non-factorizable unitary transformations do not have an efficient local realistic description, in the sense that it would demand an exponentially large number of hidden-variables \cite{Caves}. Although the correlations between the relative orientations of the spins in such experiments can be explained by means of a local realistic theory \cite{Caves}, wave-particle duality cannot be explained in the same terms.

In order to describe duality in classical terms, an infinite set of hidden-variables would be necessary, even in the presence of high levels of white noise. By Occam's razor we choose the simpler description given by quantum mechanics. The present results show that the same level of hidden-variable conspiracy reached in Ref. \cite{Terno} is also achieved in an entanglement-free scenario. This fact makes the result both sharper and more experimentally relevant, particularly in light of the discussions presented in Refs. \cite{Popescu,Kaiser}, which pointed out a possible loophole in the experiment. It is important to note that, for $\varepsilon = 0$, Eq. (\ref{eq1}) indeed has a solution. But, in this case, the action of non-factorizable unitary transformations could not be distinguished from local rotations. In other words, coherent interactions between system and ancilla could not be distinguished from non-coherent ones, leading to the indistinguishably of WL and PL photons.

In summary, we have demonstrated that the distribution given in Eq. (\ref{statmisto}) --- experimentally verified in Refs. \cite{Auccaise,Mahesh,Popescu,Kaiser,Guo}--- excludes the wave or particle behavior of any single quantum as realistic properties, even in the presence of high levels of white noise. This result implies that the wave-particle duality cannot be explained in any classical terms. To the best of our knowledge, it is the first demonstration of the non existence of a (classical) hidden-variable model in the \emph{absence} of entanglement.

\begin{acknowledgments}
The authors thank J. A. de Barros for helpful discussions about the subject of this work. It is a pleasure to acknowledge K. Modi, D. Terno and R. Ionicioiu for critically reading the manuscript. This work was supported by CAPES, CNPq, FAPESP, FAPERJ, and the Brazilian National Institute for Science and Technology of Quantum Information (INCT-IQ). LCC specially thanks for the warm hospitality of the Brazilian Centre for Physics Research (CBPF) in Rio de Janeiro, where this work was conducted. LCC, RSS and JGF also thank Moustache's pub for providing the inspiring beers.
\end{acknowledgments}

\end{document}